# Googling for Abortion: Search Engine Mediation of Abortion Accessibility in the United States

YELENA MEJOVA
ISI Foundation, Italy[1]

TATIANA GRACYK
Cleveland State University, OH, USA

RONALD E. ROBERTSON
Northeastern University, MA, USA
Stanford University, CA, USA

Among the myriad barriers to abortion access, crisis pregnancy centers (CPCs) pose an additional difficulty by targeting women with unexpected or "crisis" pregnancies in order to dissuade them from the procedure. Web search engines may prove to be another barrier, being in a powerful position to direct their users to health information, and above all, health services. In this study we ask, to what degree does Google Search provide quality responses to users searching for an abortion provider, specifically in terms of directing them to abortion clinics (ACs) or CPCs. To answer this question, we considered the scenario of a woman searching for abortion services online, and conducted 10 abortion-related queries from 467 locations across the United States once a week for 14 weeks. Overall, among Google's location results that feature businesses alongside a map, 79.4% were ACs, and 6.9% were CPCs. When an AC was returned, it was the closest known AC location 86.9% of the time. However, when a CPC appeared in a result set, it was the closest one to the search location 75.9% of the time. Examining correlates of AC results, we found that fewer AC results were returned for searches from poorer and rural areas, and those with TRAP laws governing AC facility and clinician requirements. We also observed that Google's performance on our queries significantly improved following a major algorithm update. These results have important implications concerning health access quality and equity, both for individual users and public health policy.

_______________________
[1] Corresponding Author: yelenamejova@acm.org

*Keywords: abortion, search engines, medical access, Google*

## Introduction

Abortion is one of the most contentious reproductive health services available in the United States. The Guttmacher Institute estimates 862,320 abortions were provided in clinical settings in the US in 2017, 95% of which were provided by abortion clinics (Jones et al., 2019). However, the number and geography of abortion clinics (ACs) fluctuate over time, as gestational limits and local laws regulating clinic staff and building codes – so-called Targeted Restrictions on Abortion Providers (TRAP) laws – pose restrictions on abortion clinics in various states around the country (Guttmacher Institute, 2020) and as funding rules continue to change (Guttmacher Institute, 2016; Merz et al., 1995).

Those seeking abortion procedures must often pass through an additional hurdle as well – finding an abortion clinic via a Web search engine. In 2015, people conducted 3.4 million searches for abortion clinics (Stephens-Davidowitz, 2016), and a small-scale survey (n = 356) from the same time period found that a large portion of women who obtained an abortion in Nebraska (45%) located their abortion clinic via online search (French et al., 2016), but little is known about the information that they might have seen during those searches. Having the power to determine "who sees what under what circumstances" (Halavais, 2018), Web search engines have become the gatekeepers not only to health knowledge stored in servers around the world, but to the very health services and businesses providing them. This development disproportionately affects women because they are more likely than men to search for health-related information on the Web (Fox & Duggan, 2013). Further complicating matters, all search engines must compete with spammers and scammers to maintain reliable links and business information (Copeland & Bindley, 2019; Metaxas & DeStefano, 2005).

Given this, the quality of search results for abortion-related queries has been the subject of intense scrutiny, especially for Google Search, the most popular search engine. As recently as September 2019, journalists documented Google returning search results for abortion related queries that not only do not provide abortion services, but are actively lobbying against them (Marty, 2018; Sherman & Uberti, 2019). In particular, "crisis pregnancy centers" (CPCs) – locations that are not medical facilities but often provide counseling, testing for sexually transmitted infections (STI), and ultrasounds – compete with abortion clinics to attract pregnant women (Kimport, 2020). CPCs often target women with unintended or "crisis" pregnancies who might be considering abortion (Bryant & Swartz, 2018), and the ads run by CPCs were once removed by Google for violating the search engine's advertising policy, which prohibits misleading, inaccurate, and deceitful ads (Hattem, 2014). Previous research finds such centers may provide misleading information in order to dissuade women from getting abortions (Ahmed, 2015), such as giving inaccurate medical information regarding the risks of abortion (Bryant et al., 2014; Bryant & Levi, 2012; Campbell, 2017). In light of this, some have judged these centers to be "legal but unethical" due to their deceptive practices and lack of patient-centered care (Bryant & Swartz, 2018).

While CPCs add to the myriad barriers women face when accessing abortion services in the US, there is also a large absence of abortion clinics. In 2014, 90% of US counties lacked an abortion clinic (Jones & Jerman, 2017), and the overall number of abortion providers declined by 5% from 2014 to 2017 (Jones et al., 2019). Beyond limited numbers of clinics, other major barriers to abortion include financial issues, state or clinic restrictions, system navigation issues, and travel-related logistical issues. These can result in delays in care, negative mental health impacts, and even considerations of self-inducing abortions (Jerman et al., 2017). These barriers also often interact: for example, those needing financial assistance must often travel the furthest for the procedure (Ely et al., 2017). The lack of a comprehensive listing of clinics provides yet another obstacle, creating challenges for physicians seeking to provide quality referrals (Yanow, 2009). Web search engines are in a unique position to alleviate this problem by providing an exhaustive, personalized view of the available service providers.

Thus, our research examines the role Web search plays in abortion access. In particular, we consider the scenario of a woman searching for abortion services using Google Search. We aim to answer the following research questions:

**RQ1.** What is the quality of the search results in terms of (i) the relative number of abortion clinic versus crisis pregnancy center locations returned, (ii) the presentation of the results returned, and (iii) the degree to which the results returned reflect real-world abortion access?

**RQ2.** What is the relationship between abortion clinic search results and demographic characteristics of query locations in terms of population, urbanization, income, and political leaning, as well as laws applicable to abortion clinics?

In posing these research questions, we have attempted to quantify the extent to which Google provides a helpful "view" of the health resources available to those seeking abortion services. This is distinct from prior work, which used search engines to locate abortion clinics across the United States (Cartwright et al., 2018), in that we consider a localized view from the perspective of a hypothetical search engine user located in a particular US county. Similar assessments of search result quality have been recently performed for searches related to mental health, suicide (Borge et al., 2021; Haim et al., 2017), and urolithiasis (kidney, bladder or urethra stones) (Chang et al., 2016). A general algorithmic framework for assessing the "quality" of health-related web pages has also been proposed (Oroszlányová et al., 2018), and researchers with access to search engine logs have focused on queries to understand user information needs (Abebe et al., 2019). The closest existing study queried three major search engines for abortion services in 68 major US cities in 2016-2017 and reported the proportion of web, ad, and location (map) results that either facilitated, did not facilitate, or hindered abortion access (Dodge et al., 2018). In our work, we consider the quality and stability of results over time and from a variety of locations across the US, while focusing in particular on the distinction between search results which feature facilities offering abortion services versus CPCs, as they may compete with the former for the user's attention.

## Methods

Below, we describe how we operationalized the information needs related to seeking an abortion. Using 10 abortion-related search queries, we conducted searches from 467 different locations across the US over 14 consecutive weeks. After establishing which location results that Google returned were CPCs or ACs, we examined their distribution across the states, compared them to known locations of ACs, and contextualized these results with local variables concerning demographics and TRAP laws, as well as search-specific variables including time of search and result presentation.

### *Location Selection*

Location is a factor known to affect the composition of Google's search results (Ballatore et al., 2017; Kliman-Silver et al., 2015). In order to use this localization of search results within the context of our research questions about the accessibility of abortion information, we selected a diverse set of counties across the United States. Besides covering all 50 states, we aimed to ensure variability in terms of (a) population of the city, (b) socioeconomic status as measured by median household income, (c) extent of population living in rural areas, and (d) overall political leaning of the county's residents. To capture (a)-(c), we used county-level US Census estimates from 2016, and for political leaning we used the proportion of residents who voted for the Republican candidate in the 2016 US Presidential Election. Each county was assigned a binarized indicator for each of the four attributes (below or above the median), and a stratified random sampling was performed over these attributes, plus the state (Neyman, 1992). This procedure resulted in 467 locations.

*Query Selection*

We based our selection of queries that women may use when searching for an abortion clinic on the broad root query "abortion." We focus in particular on women who have made the decision to seek an abortion at an abortion clinic, as opposed to women who are pregnant and still weighing their options. For instance, we do not focus on those who are interested in details about the abortion procedure, its alternatives, costs, or anything else that may signal they have not yet made up their mind. We then expanded on the root query of "abortion" by using it as an input for Google's Autocomplete, People Also Ask, and Related Searches features. These features can be useful for query expansion because their search suggestions represent what Google infers, based on internal and otherwise inaccessible data on what real users are searching, to be relevant queries and questions for a given root query (Robertson et al., 2019).

We then assessed the relevance of the query suggestions that emerged during this exploratory process by conferring with a subject matter expert (a PhD in Applied Philosophy specializing in Bioethics). At the end of this process, we selected a total of 20 queries that a person looking for abortion services may use, and placed these queries into one of five categories: (1) general: "abortion", "pregnancy termination", "abortion clinic" (+"near me"), "abortion center" (+"near me"), (2) informational: "kinds of abortion", "types of abortion", "how does abortion work", (3) safety: "is abortion safe", "is abortion dangerous", (4) cost: "free abortion", "abortion cost", "does insurance cover abortion", and (5) legality: "is abortion legal" (+"in my state"), "is abortion illegal" (+"in my state"). Two of the queries we selected – "abortion clinic [location]" and "abortion center [location]" – were dynamic, and included the name of the county they were being searched from (e.g. "abortion clinic Smartsville California"). Upon examining our data, we discovered that some queries never produced any location search results. Given that these results are the focus of our study, we filtered out those queries, leaving us with a final set of 10 queries. These queries were: "abortion," "free abortion," "abortion cost," "pregnancy termination," "abortion center," "abortion clinic," "abortion center near me," "abortion clinic near me," "abortion center [location]," and "abortion clinic [location]."

### *Conducting Searches*

To conduct our searches, we used an algorithm auditing approach to investigating black-box systems (Metaxa et al., 2021) with an open source python library called WebSearcher (Robertson & Wilson, 2020).[2] This library provides functions for sending Google Search a customizable HTTP request and parsing the Search Engine Results Pages (SERPs) that it returns. We modified the requests we sent to reflect what users might see if they had conducted our queries from a desktop computer, with a modern operating system and web browser, and while logged out of any online accounts. Although we study location-based personalization, we explicitly do not study personalization at the individual level, because our searches were not connected to any accounts or online identifiers other than a persistent IP address and User-Agent.

WebSearcher also enables the geolocation of searches by modifying a URL parameter to include an encoded location name. More specifically, this parameter enables one to search from any location that has a "Canonical Name" in Google's geotargets for advertisers.[3] We verified that changing this parameter geolocated the results that Google returned by conducting several queries (often involving food, e.g. "pizza") from several locations, manually examining any location results present in Google's SERPs, and confirming that the results were for the location we had selected.

Using our 10 queries and 467 locations, we set up a script that used WebSearcher to periodically search each query from each location. The script was automatically executed every Monday morning for 14 weeks of our study. Conducting multiple searches using the same queries at different points in time helps to measure and account for temporal differences in the search results (e.g. the news articles returned can frequently change), enabling us to detect changes and adding to the temporal validity of our study (Munger, 2019).

---

[2] https://github.com/gitronald/WebSearcher
[3] https://developers.google.com/adwords/api/docs/appendix/geotargeting

For each SERP we collected, we used WebSearcher's parser to systematically extract details from the various components that compose a modern SERP. These components included standard results, ads, knowledge boxes, images, news cards, and location results, among others (Robertson et al., 2018). For each component in a SERP, the parser noted the type of component it was, the rank at which it appeared, and extracted its text, links, and other information. In this study we focus on the location components – which contain up to 3 map locations of businesses – and typically include a business name, address, and possibly a phone and website. We leave the exploration of the remaining result types for future work.

### *Clinic Listings*

In order to compare the search results returned by Google to existing abortion clinics, we obtained the latest information on the number of abortion clinics (ACs) and crisis pregnancy centers (CPCs) across the United States.

Due to the sensitive nature of this information, as abortion clinics are often targeted by protesters and incidents of violence and deaths have occurred, there are no publicly available complete listings of ACs. However, we were able to obtain three sources of information. First, from the National Abortion Federation (NAF) (https://prochoice.org/), a professional association of abortion providers in North America. On their website, the NFA provided a list of 294 clinics in the United States which we downloaded in June 2019. Second, from the Guttmacher Institute, a research and policy organization which produces reports on the state of contraceptive and abortion services. Although they do not provide the listing of clinics they have collected, they provided a state-aggregated count from 2017 (Jones et al., 2019). The clinics were collected through "Web-based searches, media reports, and reviews of directories of organizations and associations that work with abortion-providing facilities," as well as "a national survey of obstetrician-gynecologists" and "state health department data." In total, the Guttmacher list provided 808 clinics. The third, and most complete dataset, came from Cartwright et al., which used Google, Yahoo, and Bing's search engines to compile a complete list of clinics in the United States (Cartwright et al., 2018). This list of clinics, manually updated in 2019, was made available to us for research purposes by Advancing New Standards in Reproductive Health (ANSIRH) upon our signing of a confidentiality agreement. Note that, considering per-state clinic statistics, the NAF listing correlates only loosely with the other two (Pearson correlation coefficient at about $r = .62$), while the Guttmacher and ANSIRH lists are strongly correlated ($r = .992$), corroborating their version of "ground truth" in terms of the per-state abortion clinic distribution.

Unlike ACs, extensive lists of CPCs can be found via several sources. We used three: (1) CareNet, a nonprofit pro-life organization that has 1,703 locations listed on its website (care-net.org), (2) Heartbeat International, a network of "pro-life pregnancy resource centers" which has 4,090 locations listed on its website (heartbeatinternational.org), and (3) a list of clinics not providing abortion services at "exposefakeclinics.com," manually checked for accuracy by the authors. The first two lists we obtained in June 2019, and the last in October 2019. Note at this stage the difference in prevalence of the two kinds of clinics, with the highest estimate of ACs at around 800 and of CPCs at 4,000.

### *Labeling Location Results*

We approached the task of understanding whether the returned locations were in fact abortion clinics by first de-duplicating the search results, then matching as many as possible to the available clinic lists, and finally coding the rest manually. For each location result we extracted several fields, including the title of the business, URL, and contact (which may contain the address or phone number), with the latter two possibly not being present. After de-duplication across all queries, locations, and the 14 days of data collection, we obtained 4,388 results (note that this set still had potential near-duplicates due to address formatting, etc. which were hand-checked at a later stage).

These results were then compared to the lists of ACs and CPCs, with exact matching on phone numbers, when available, and approximate string matching for business titles and addresses. Each potential match was manually examined to make sure the title of the business and location matched. The rest of the results were also manually examined and annotated by the authors with the following guidelines:

- a clinic that provides abortion services (including the abortion pill – "medication abortion" – not just the morning after pill) was labeled as an abortion clinic (AC),
- a service center for pregnant women that discusses abortion services, but does not refer for nor provide abortion services nor provide any other medical services (besides ultrasounds and STI testing) was labeled as a crisis pregnancy center (CPC) (as defined by the National Abortion Federation),
- any other clinic or hospital which does not provide abortion services, but does provide medical services (other than ultrasounds and STI testing) was labeled simply as "clinic,"
- and all other results were labeled as "other."

Note that the distinction between an AC and a CPC label is often a very small but important one – whether abortion services are in fact provided. Numerous CPC websites provide a detailed description of abortion services, to only at the bottom of the abortion description page, or on another page altogether, declare that abortion services are not provided (see Discussion section on potential attempts of such pages to use "search engine optimization" to get a higher chance of being shown in a SERP).

Yet another important distinction is between CPC and clinic locations, which are mainly distinguished by whether the location provides services other than an ultrasound and STI testing. For instance, OB/GYN or midwife services cater to pregnant women, but unlike CPCs, provide medical services. Note that clinics may provide referrals to ACs, but this varies across the country, and in some parts of the US referrals have been a target for legislation seeking to limit access to abortion (Zurek & O'Donnell, 2019). Because of the sensitivity of the task, an annotator agreement exercise was performed on 25 randomly chosen locations wherein all three annotators independently applied one of the four above labels to the results, which upon later comparison showed near perfect agreement. Note, however, that there were several results where individual annotators had to consult each other, and the labels for these were decided by a majority vote.

After manual de-duplication and removal of "other" results, we produced a list of 1,569 unique location results – 553 ACs, 369 CPCs, and 647 clinics. The remaining 745 "other" results varied dramatically in topic, including other medical services (dentistry, substance addiction counseling, veterinary clinics), government websites (of cities, counties, or states), various stores (selling clothing, jewelry, etc.), penitentiaries, and churches (16% of all "other" results), among others.

During labeling, we also noted the city and state of the results, taking the address from the website to which the result linked. If we couldn't find one there, we used the address Google provided on the SERP. We then used Google's Geocoding API to map the addresses to GPS coordinates, which could then be compared to the origins of queries using geodesic distance (the shortest distance on the surface of an ellipsoidal model of the earth), in miles.

We finished the processing of the results by matching them with relevant state-level and county-level (when possible) information such as the number of AC and CPC locations known in other resources, demographic data, and abortion-specific legislation (Guttmacher Institute, 2020). The latter encompasses restrictions on the procedures performed at the clinics, facility dimensions and position, and certification of the clinical staff.

### Data Availability

Because of the sensitive nature of the data involved in this study, we employ the data sharing approach devised by Advancing New Standards in Reproductive Health (ANSIRH): the data will be available to the research community upon request and signing of a confidentiality agreement. In particular, the parsed location search results over 14 query sessions, their manual annotation, address and geo-location will be made available, in order to ensure reproducibility and continuation of this line of research.

### Location Type Prevalence

In order to gauge the prevalence of the different search result types (AC, CPC, or clinic), we consider results returned for all of the 14 query sessions as one large session, and report the percentage of results observed for each type. For state-specific analyses, we averaged the number of different types of results returned over all locations in that state. We then used Spearman's Rank correlation to compare these figures to the number of ACs in each state, as determined by the ANSIRH database (described above). Due to the right-skewed distribution of ACs per state (33 states have 10 or fewer ACs, while California has 152), we used the nonparametric Spearman Rank correlation here rather than the Pearson correlation. We also used the two-sided T-test to compare the number of ACs in states where CPCs were and were not returned. Finally, we plotted the distribution of result types per state, as well as the average number of ACs or CPCs returned in each state geographically.

### Distance to Query Origin

We computed the distance of the returned results to the query origin in miles, using the GPS coordinates of their addresses. We then plotted the distance to the origin by result type. We also computed the distance of the closest result returned in a session (one which may potentially attract the user most). Finally, we compared the closest returned AC to the best-case-known-scenario, or an "oracle", which is the closest AC as indicated by the ANSIRH database. When we plotted the closest result AC and closest known AC, we checked to see whether the points lay on the diagonal, that is, whether the closest known AC was returned by the search engine. If not, we noted any discrepancies between Google's results and known ACs around the query point.

### Comparison to Demographics and TRAP Laws

Next, we asked whether the number of AC or CPC results returned by Google were in some way related to the demographics of the location from which the query was issued. In particular, we modeled the total number of (non-unique) AC or CPC results returned for a search (although results were similar when modeling the number of distinct businesses returned) using the pooled linear regression model, treating the 14 query sessions as panels.

As a baseline, we used the log of the number of ACs in each state from the ANSIRH database. We also considered adding a similar count of CPCs per state to the model, but this increased multicollinearity. The per-county demographic variables of median household income, percent rural, and percent voting for the Republican US Presidential candidate in 2016 were z-normalized for ease of comparison. We excluded population size due to a high correlation with our rural measure (Pearson r = -.44). Additionally, we included information about the per-state TRAP laws as 5 components obtained via Principal Component Analysis (PCA), employing Principal component regression (Cui et al., 2011; Massy, 1965). These 5 components explained 80% of variance and did not contribute to multicollinearity of the predictors. The final model had maximum Variance Inflation Factor (VIF) of 1.12, indicating low multicollinearity. The *P*-values were computed using a permutation test with n = 1000.

### Result Stability Over Time

Since our data was obtained over 14 weeks, and search results are known to change over time – due to the thousands of minor updates that Google makes to its search engine every year,[4] as well as changes to the online landscape – we examined the stability of the location results we collected over time. To do so, we plotted the proportion of the four location types that we observed (aggregated across all locations and queries), by the date that they were collected. For each consecutive pair of data collection dates, we ran a $\chi^2$ test in order to test for significant differences in the distribution of location types. To better understand any significant changes that emerged from this exploration, we also examined this data in relation to the date of a publicly announced major algorithm update, in which Google incorporated a new technology known as BERT (Bidirectional Encoder Representations from Transformers) to better understand users' queries (Nayak, 2019).

---

[4] https://moz.com/google-algorithm-change

### *Result Presentation*

We further summarized several contextual elements of the result presentation, mainly the ranking of the results, the number of user reviewers and their aggregated score, the classification of the business, and the business title. More specifically, we present the proportion of time each result type was presented at a ranking, because a location result's ranking within the search result page signifies the relevance of the result, and search rankings can steer users' attention and influence their opinions (Epstein & Robertson, 2015; Pan et al., 2007). Google also allows users to leave reviews and ratings of businesses, creating crowd-sourced social signals for quality. To capture this, we plotted the distributions of the number of reviews and the average ratings for each result type. In addition to this, businesses listed in Google's Search rankings are often presented with a category label. When business owners add their listing to Google, they are able to search for and select a pre-existing category from a drop-down list of options provided by Google, or input a more specific category of their own choosing.[5] However, once published, these categories can also be edited by users (Bonelli, 2018), adding an adversarial edge to the way businesses are categorized in Google Search. Our results present the instances when the categories on Google's search results do not match the labels in our data. Finally, we examined the titles of the locations (business names), as they are the most informative piece of information available to the user. To better compare these titles, we made them lowercase, tokenized them, and removed stopwords as well as any location-specific words associated with the origin of the query. Our results thus present both the most used terms by ACs and CPCs, as well as terms most distinct to each group computed via the difference of probabilities.

---

[5] https://support.google.com/business/answer/2911778/

## Results

### *Location Result Prevalence by Query*

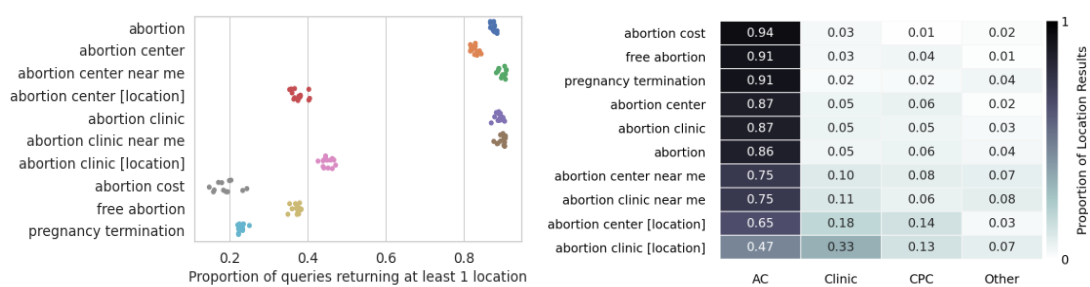

**Figure 1. (left) Average proportion of queries resulting in at least one location result, each point represents a query session. Queries with [location] were localized using the name of a city. Points are randomly jittered in y axis for visibility. (right) Proportion of location types by query. Rows ordered by proportion of ACs returned.**

  To measure the variability of results across time, we ran a set of 10 carefully selected queries once a week for 14 weeks between October 2019 and January 2020 (see Methods for query selection details). We refer to each day of data collection as a "query session." In total, 70.4% of the search pages we collected across all query sessions contained at least one location result, although this distribution was not uniform across query types. Figure 1 (left) shows the proportion of responses for each query that returned at least one location result, with each dot representing a query session, such that the variability across sessions is visible. Note that the proportion of results does not vary significantly within the 14 sessions, except perhaps for the "abortion cost" query. The highest proportion of results with locations were returned for "abortion center near me," "abortion clinic near me," as well as "abortion clinic" and simply "abortion." Comparatively fewer results were returned for "pregnancy termination," showing the importance of terminology selection when searching for abortion procedures. We also found that our localization of the query, by appending the location name, produced fewer results than their non-localized versions. For instance, 89.2% of searches for "abortion clinic" returned a location result, but only 45.1% of searches for "abortion clinic [location]" did so as well. A similar effect can be seen for "abortion center" and "abortion center [location]" (Figure 1).

### Prevalence of AC and CPC Results

The majority of results (79.4%, n = 91,095) were ACs, followed by other clinics (9.2%, n = 10,528), and CPCs (6.9%, n = 7,891). When examining each query separately in Figure 1 (right), we found that, for queries with location keywords, the prevalence of ACs was lower while the prevalence of CPCs and clinics was higher. However, this distribution was quite heterogeneous among the states, as shown in Figure 2, where black diamonds show the number of ACs in state (right y-axis). Note that not all queries produced location search results, and that a maximum of 3 location search results appeared in a single search page.

Washington DC, New Jersey, and Connecticut returned the most AC results, while Montana, North Dakota, and Alaska returned the fewest. The geographical distributions of results that were AC (left) and CPC (right) in each state are plotted in Figure 3. It shows that the coasts returned larger numbers of AC results and fewer CPC results than inland. Only one query location returned no AC results: Joplin, MO, instead returning CPC results 33% of the time (the rest being clinics). Conversely, 197 locations never returned a CPC result.

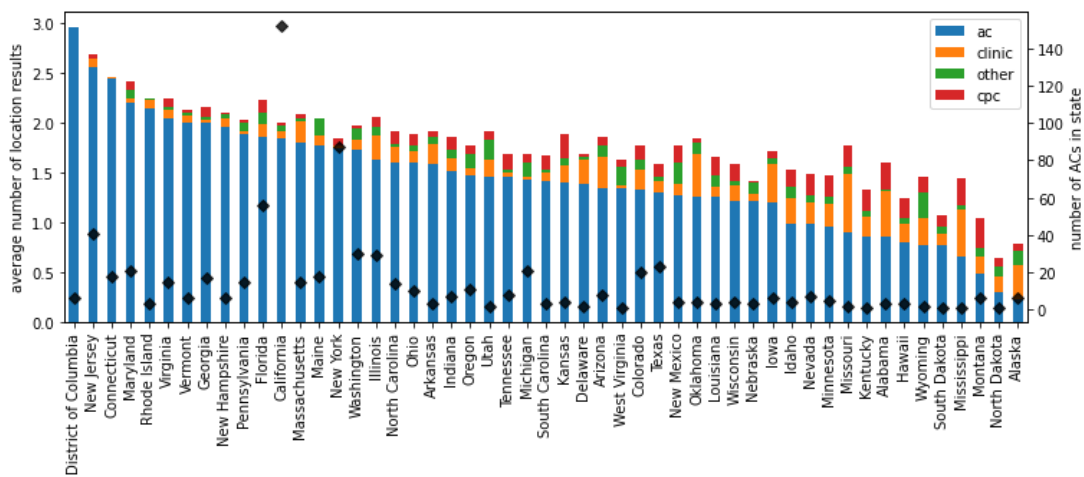

**Figure 2. Average number of location results with different labels per query, per state (bars, left y-axis) and number of ACs in each state (black diamonds, right y-axis).**

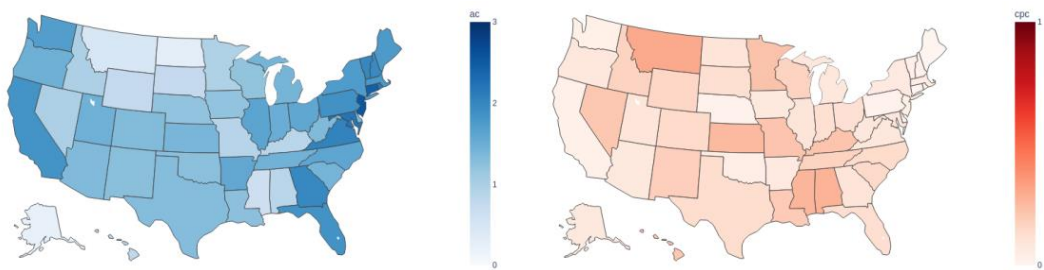

**Figure 3. Average number of results that are ACs (left) and CPCs (right) per query. While each has a maximum of 3 possible results, we plot the average number of CPC results returned on a scale of 0 to 1 to better highlight the distribution, because the max value was 0.30 in Montana.**

Comparing the known number of ACs (determined using the ANSIRH database) to the average number of AC results returned per search in each state, we found a Spearman Rank correlation of ρ = .624 (P < .001). We also checked whether CPC results were returned for query origins that had few ACs in the vicinity. Indeed, we found that locations with CPC results were more common in states with lower numbers of ACs. States in which CPCs appeared in the results had on average 14.6 ACs, and those that did not had on average 20.5 ACs (significant at P = .03 using a two-sided T-test). This suggests that a higher prevalence of ACs in a state makes it more likely that Google will return location results for ACs, and this may be pushing any CPC results out of the top three rank positions. Interestingly, the number of CPCs in the state correlates poorly with the number of CPC results in the SERP (Spearman's $\rho = 0.15$, P = 0.293), as well as the number of AC results (Spearman's $\rho = 0.22$, P = 0.128), possibly indicating that the number of CPCs in a state is not as important of an indicator as the (much smaller) availability of ACs, when the web user is searching for abortion providers. Further, while the number of ACs in a state appears to matter for whether a CPC result was returned, we found no relationship between whether a CPC result was returned and the distance between the query origin to the closest AC result.

### *Distance to Query Origin*

Figure 4 (left) summarizes the distances of AC or CPC results to the origin of the query. From the insert, which shows the raw number of results, we can again observe the majority of results being identified as ACs. However, considering the proportional distribution in the main figure, we find that most of the CPC results that were returned were much closer to the query origin. Specifically, the mean distance for an AC result was 73 miles (median 37), and for a CPC result was 66 miles (median 8), with the difference in medians signifying that while both distance distributions were skewed, they were skewed more significantly for CPC results.

Further, in the cases when both AC and CPC locations were returned for a query, CPCs were the closest result 75.9% (4,451 out of 5,872) of the time, meaning that if a CPC was present in the result set, it was likely be the closest result to the query origin. This could be due to the fact that there are significantly more CPCs than ACs (judging from the sources we used), with roughly 4,000 CPCs and approximately 800 ACs in the United States.

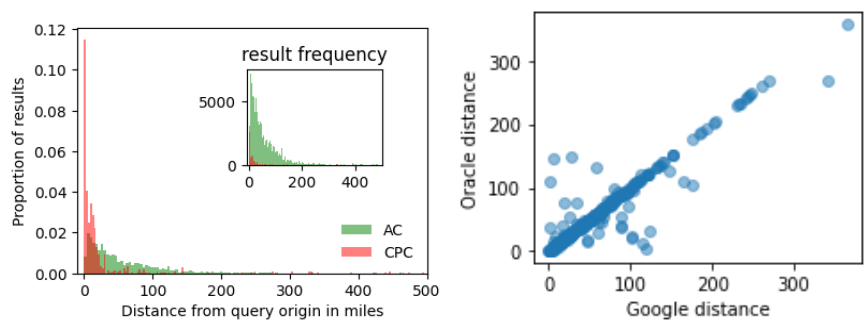

**Figure 4. (Left) Proportion of results at a distance (in miles) from the origin of the search, with raw number of results in the insert. (Right) Distance to closest AC for Google results vs. Oracle, per query origin.**

Finally, we checked to see how well the closest returned AC corresponded to ACs in the real world, as provided by the ANSIRH database. For each query origin, we computed the distance to the closest AC in the database, referred to as an "oracle" estimate. Figure 4 (right) plots the distance to the closest Google location vs. to the oracle location for each query origin location over all query sessions (that is, the closest AC across all sessions). We found that most fell on the diagonal, with 90.5% of returned results within 5 miles of the oracle AC. Where the points are not on the diagonal, it is because the Google result was closer in 18 cases and the oracle was closer in 26 cases. When we manually examined the instances where Google found closer alternatives, we found that these were either not in the ANSIRH database or were marked as not providing abortions, despite advertising the procedure on their website. Because we rely on the business websites for the labels, we will leave updating the database with the latest information to future work. If we constrain our examination to the closest AC within a query session (that is, what a user would have seen that day if they searched our queries), the closest AC is within 5 miles of the oracle in slightly fewer cases (86.9% of the time) with only 2.5% of cases having their closest AC result over 50 miles away from the oracle. In all, we found that Google's search engine provided the closest possible ACs in the majority of the cases.

### *Demographics & Laws*

Next, we asked how many times a woman is likely to encounter an AC in the search results when conducting a query on a particular day from a particular location. Table 1 shows the β coefficients and the corresponding P-values of two models – one for number of AC results (left) and another for CPC results (right) – for demographic variables and TRAP laws, after distributing component coefficients via their PCA loadings. For brevity, we omit the dummy variables used for each of the 14 query sessions.

We found that the number of AC results returned is strongly associated with the number of ACs in the state, indicating that the search engine generally reflects the availability of abortion services in each state. The strongest demographic effect was income, which had a positive association with the number of ACs returned, followed by rural and Republican, both having negative associations. All coefficients reverse when we model the number of CPC results, except for rural, where the association remains negative. Considering specific law requirements (TRAP Laws), the strongest negative effect we found on returned ACs was for requirements that clinicians have hospital privileges, followed by requirements for facilities (distance from a hospital). These per-law AC model coefficients remained stable when other numbers of PCA components were used, with a maximum variability of .044 for 4 or 6 components, indicating that our results are robust to such adjustments. Finally, we note that a better fit is achieved for modeling the number of ACs returned ($R^2$ = .254, n = 463) than CPCs ($R^2$ = .125, n = 463).

**Table 1. Pooled linear regression models predicting number of AC results and CPC results at a location in a query session, showing the β coefficients and associated P-values. TRAP law coefficients aggregated via 5 PCA components. Dummy session effects omitted.**

| | AC | | CPC | |
|---|---|---|---|---|
| | **β** | **P** | **β** | **P** |
| constant | 8.765 | <.001 | 1.837 | <.001 |
| AC in state ANSIRH (log) | 2.406 | <.001 | -0.296 | <.001 |
| republican$_z$ | -0.232 | <.001 | 0.168 | <.001 |
| rural$_z$ | -0.411 | <.001 | -0.491 | <.001 |
| income$_z$ | 1.471 | <.001 | -0.203 | <.001 |
| **TRAP Laws** | | | | |
| clinician hospital privileges | -1.351 | <.001 | 0.197 | <.001 |
| facility hospital max distance | -0.896 | <.001 | -0.177 | <.001 |
| facility corridor width | -0.466 | .02 | 0.284 | <.001 |
| clinician OBGYN certification | -0.290 | .004 | 0.160 | <.001 |
| facility hospital transfer agreement | -0.051 | .02 | 0.039 | .004 |
| service pill | -0.044 | .02 | -0.141 | .01 |
| service surgical outpatient | 0.064 | .001 | -0.029 | .02 |
| facility surgical standards | 0.126 | <.001 | -0.037 | <.001 |
| service surgical private doctor | 0.355 | <.001 | -0.146 | <.001 |
| facility procedure room size | 0.944 | <.001 | -0.455 | <.001 |
| | $R^2 = 0.254$ | | $R^2 = 0.125$ | |
| | $n = 463$ | | $n = 463$ | |

***Result Stability Over Time***

Next, we examined the stability of the results over the 14 query sessions, as shown in Figure 5. Although the proportion of locations returned by Google was relatively stable across our first four sessions, we observed a 5% increase in the proportion of AC results between October 28 and November 11, 2019 ($\chi^2$ (3, n = 16,254) P < .001). These changes could potentially be due to changes in the search index or algorithm, such as Google's BERT algorithm update, intended to better "understand" users' natural language queries, which was officially introduced on October 25, 2019 (Nayak, 2019). This hypothesis is somewhat supported by the observation that performance improved more for locations with fewer ACs in their state (split on median AC normalized per person). States with fewer ACs received an additional 1.27 AC and 0.37 fewer CPC results (both P < .001), whereas those having more ACs received only an additional 0.73 AC and 0.06 fewer CPC results, on average (only AC change significant at P < .001).

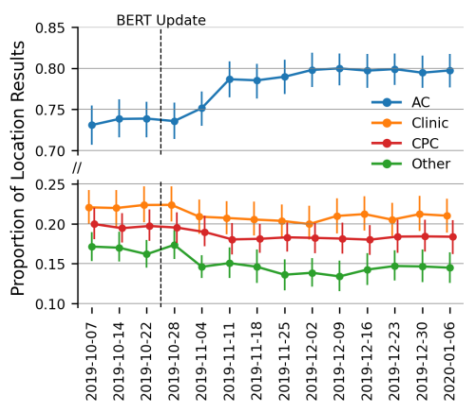

**Figure 5. Proportion of locations over data collection dates (with 95% confidence intervals). The proportion of AC's returned in Google Search increased by over 5% after the BERT update** (Nayak, 2019) **(dotted line).**

### *Result Presentation*

In Figure 6 (left) we show the proportion of times each kind of result was shown in each rank given a location type (out of 3 possible). We found that Google's search engine generally ranked AC results in the top two ranks, while CPC results generally occupied lower ranks, and other results were ranked the lowest.

The distributions of the number of reviews for each location type in our dataset are also presented in Figure 6 (center), showing that ACs received the most reviews (median of 22), followed by CPCs (9), clinics (8), and other (5). Figure 6 (right) shows the distributions of average ratings for the locations in the four categories. We observed the lowest median rating to be for ACs at 3.7, followed by clinics at 4.0, CPCs at 4.3, and 4.6 for others. The differences in ratings for ACs vs. CPCs and clinics vs. CPCs are significant at $P < .001$ using Mann-Whitney U test (statistic = 38475), but not for ACs vs. clinics ($P = .075$). Unfortunately, the results pages do not provide any further information about the reviews, and we leave an in-depth examination of this data for future research.

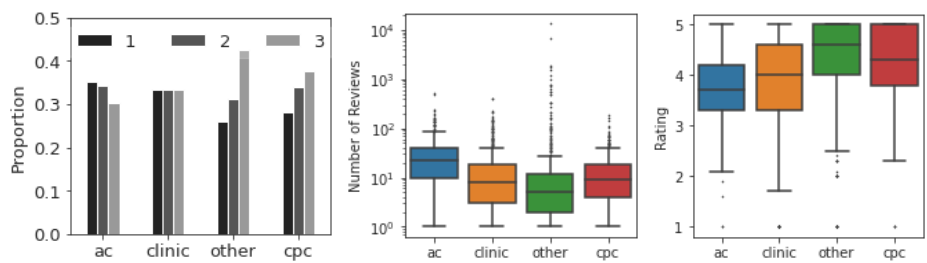

**Figure 6. Proportion of time result of a different category is shown at a rank (left), distributions of the number of reviews (center) and average rating (right) of locations in different categories.**

Examining the business category labels that Google often presents alongside location results, we found that the one most associated with the results we labeled as AC were "Abortion clinic" (n = 516, 87%), followed by "Medical Center" (n = 24, 4%), "Family planning center" (n = 18, 3%) and "Women's health clinic" (n = 15, 2.5%). For CPC results, the most common categories were "Pregnancy care center" (n = 199, 54%), "Women's health clinic" (n = 57, 15%), and "Medical clinic" (n = 30, 8%). Given that our own labeling criteria for CPCs included that they do not provide gynecology or obstetrics services, we manually checked 30 locations with websites that we labeled as CPCs, but which were categorized as "Medical clinic" on Google. We found that none of these CPCs provided gynecological services: only 14 provided limited ultrasound services and STI testing, with the rest providing one or the other, or no medical services at all. Also, four of the locations we identified as CPCs were inaccurately categorized as "Abortion Clinic" (1% of total CPC results) on Google. Finally, we found 217 locations changed their categories in the span of our data collection period, 87% (n = 188) of which were ACs, and out of these, 75% (n = 141) of the time they changed their label from "Abortion Clinic" to something else. Most of the switches happened in October 2019, potentially in connection to the improved performance that we observed (Figure 5). The figures above are computed on the latest categories in our dataset.

Further, we also examined the words present in the titles of the locations (business names). We observed that besides "planned" and "parenthood" being more prevalent in ACs, many words were used similarly, with the top 11 including "center," "clinic," "women," "medical," and "abortion" (the last was used 1.2% by AC and 1.7% by CPC locations). To clarify the difference in word usage, we subtracted the probability of seeing a word used by one group from another, and found that ACs emphasize medical aspects such as "health," "reproductive," "surgical," and "gyn," while CPCs emphasize emotional aspects such as "care," "life," "hope," "choices," and "help" (words, again, from the top 10 words by normalized probability).

## Discussion

Overall, Google's location results for abortion-related queries were relatively accurate, relevant, and improved over the course of our study. On average, 79.4% (n = 91,095) of the results we observed were ACs, contrasted with only 6.9% (n = 7,891) CPCs, a marked improvement over the previously reported 21.7% CPCs (Dodge et al., 2018). The AC results were also more likely to appear at a higher rank in Google's local results than CPC results (Figure 6), and in 86.9% (n = 5,559) of query sessions, the closest AC returned was within 5 miles of the known AC location.

However, this performance was uneven across the locations we examined – the regression analysis we conducted shows that searches conducted from states with fewer known ACs were likely to return fewer AC results, and, alarmingly, more CPC results. This indicates that Google's search results reflect existing inequalities in abortion access: when abortion access is restricted within a state due to a limited number of existing ACs, women searching for abortion services will also see fewer AC and more CPC search results returned. When CPCs were present in the result set, they were also the closest result to the query origin 75.9% of the time. Additional information provided on Google's SERPs via user-submitted reviews show that ACs have the lowest ratings and the highest number of reviews among the location types we examined, suggesting that these locations receive a greater amount of attention from users (or potential adversarial bot or spam ratings). Further, four locations we identified as CPCs were classified as "Abortion Clinic," and 30 CPCs as "Medical Clinic" despite providing limited medical services. Thus, the quality of results can still be improved, especially for underserved locations, by providing more ACs (even if they are further away), fewer CPCs, and ensuring the integrity of their listed categories.

Concerning the number of ACs returned for queries in various locations across the country, we found fewer AC results returned for those living in rural areas, and more AC results returned for those with a higher median income. With respect to political leaning, as measured using 2016 US Presidential votes, we found a smaller but significant relationship, with fewer AC results and more CPC results being shown in locations that voted for Trump. Lastly, we found a negative relationship between the number of ACs returned and TRAP laws that restrict facility buildings and create hospital privilege requirements. Thus, we find that Google reflects several of the current demographic and policy inequalities in abortion access, a finding which echoes prior work on inequalities in how minority groups are represented on Google Search, and adds to calls for more work on this powerful medium (Noble, 2018).

### *User Implications*

Our work has several important implications for users. First, we found that query wording can have a substantial effect on whether location results were returned, for instance, including "near me" or a location name in the query actually decreased the chances of receiving an AC, possibly because these "dilute" the importance of "abortion" as a keyword, as earlier studies have suggested (Dodge et al., 2018). This may present a real obstacle to finding proper medical care, as it has been shown that young adults have difficulty finding the correct keywords to find accurate information about emergency contraception (Hargittai & Young, 2012). More research is needed to understand how someone might formulate a query specific to finding abortion services, perhaps through ethnographic or survey-based studies (Mustafaraj et al., 2020; Trielli & Diakopoulos, 2020; Tripodi, 2018) , or through data sharing agreements with search engines, but it is clear that promoting digital literacy may be an effective way to promote access to quality information and increase healthcare utilization.

Second, when the search results were returned, they were likely to reflect the availability of ACs in the state of the query origin, as well as the demographic inequalities that especially disadvantage those searching from lower-income and less populated areas – inequalities which have long been observed (Fried, 2000). The fact that the number of results varies widely between the states means that some users will not be presented with all available alternatives, such as further-away or out-of-state ACs, propagating the existing unawareness of all available services (Yanow, 2009). This may be due to distance restrictions the search engine puts on location results. In that case, we would encourage Google to broaden the selection criteria in order to provide alternative ACs, even if they are not nearby or are across state lines, when they are most relevant to what a user is searching for. This will hopefully diminish the number of CPCs returned, even though they may be closer due to being more abundant. Recall that one location, Joplin, Missouri, returned no AC results, even though an AC was available about 70 miles away in the city of Fayetteville, AR. Thus, users may want to search for clinics in other locations, to make sure they get adequate coverage of alternatives.

Third, misclassification of CPCs as "Medical Clinics" may provide not only false information to search engine users, but may be against some consumer protection laws, which aim to ensure that businesses do not practice medicine without a proper license (Campbell, 2017). The extent to which such laws may apply to the information search engines provide in their results is unclear and needs further scrutiny by legal scholars, but the problems of keyword spamming and "Google bombing" — instances of web page designers tricking the search engine into ranking their content higher — are not new, and our findings warrant increased attention from the search engine on how their location labels are being abused to mislead users (Bar-Ilan, 2007; Gillespie, 2010, 2017; Grimmelmann, 2008; Introna & Nissenbaum, 2000). Until then, the onus is on users to closely examine a business's website and reviews to determine whether they provide the advertised services.

During the manual annotation, we noticed many CPC websites using language tailored to women seeking abortions (such as providing detailed descriptions of abortion procedures, offering "confidential abortion consultations" or discussing the cost of abortions) while disclaimers about not providing or referring to abortion services were often only displayed in fine print at the bottom of the webpages where they are unlikely to be noticed. Concretely, we manually checked a sample of 100 CPC websites, and out of 98 which were still accessible, 52 had a description of abortion (which were on average 773 words long) and out of those only 63% had a disclaimer that they do not perform abortion on that page. When looking at the entire website (not just the page describing abortion), 34 didn't have any explicit statement about not providing abortion services. As such detailed labeling is out of scope for this work, we urge the research community to further examine the quality of information provided on CPC websites.

However, services provided by CPCs may still be useful to women considering abortion (such as ultrasounds and STI testing), as well as a possible discussion of alternatives to the procedure, such as adoption. Thus, if the services provided by the businesses are clearly stated in the SERP, the inclusion of CPCs in the search results may even be of interest to those seeking abortion services. Note that during the manual labeling, some CPC websites were found to be upfront about their pro-life stance and about their aims. Thus, the extent to which these websites and businesses actually harm the women seeking abortions is an important question.

### *Policy Implications*

Among the explanatory variables we included in our model, TRAP laws were shown to be highly related to the number of ACs Google returned, even when controlling for demographic variables and the number of ACs in the state. In particular, requirements dealing with facility restrictions (room and corridor width, distance to a hospital, etc.) and clinician requirements (having hospital admission privileges) were negatively associated with the number of ACs returned. While a temporal analysis of the passing of such laws and the availability of ACs would better elucidate a causal relationship, we stipulate that the laws directly affect the operations of the ACs in each state (affecting corresponding Google searches). A parallel study of legislative attempts at regulating CPCs, such as California's Reproductive FACT Act (California Legislation, 2015), would reveal the efficacy of such approaches.

Alternatively, there are ongoing efforts aimed at the regulation of search engines and other online services in order to protect their users (Taddeo & Floridi, 2017). Although social media is attracting more attention in terms of regulation (Mackey & Liang, 2013), search engines are a primary tool for finding local businesses online (Google, 2014). Search engines, which are receiving increased legal attention at national and international levels for how they curate online information (Grasser, 2005; Noble, 2018; Trautman, 2017), may benefit from further engagement with public health institutions and researchers to monitor the quality of data they surface to their users. For example, the notion of a "filter bubble" (Pariser, 2011) that is often invoked in the political sphere, sometimes specifically in regard to Google Search (Robertson et al., 2018), can similarly be applied to the health domain (Holone, 2016).

The potential reinforcement of healthcare access inequality by the search engines may fall into the larger topic of algorithmic fairness, a topic discussed by Virginia Eubanks in "Automating Inequality: How High-Tech Tools Profile, Police and Punish the Poor" (Eubanks, 2017) and Cathy O'Neil in "Weapons of Math Destruction" (O'Neil, 2016). Search engines may be considered a part of decision-making, and the biases and inaccuracies built into their systems may be difficult to reveal without systematic oversight. As O'Neil puts it, such algorithms or models can be "self-perpetuating" and highly destructive. For example, a dearth of AC results in areas with already limited access to abortion results in the perception that these resources are unavailable, leading to these resources being underutilized, and subsequently down-ranked. Systematic study of these issues will hopefully identify locales, such as Joplin, Missouri, which may suffer from the perceived lack of health access, despite services being available in the vicinity. In the light of the ongoing debate over how search engines personalize content around political issues (Introna & Nissenbaum, 2000; Robertson, 2018), it would be interesting to perform the current study while logged into accounts associated with differing political stances (Hannak et al., 2013; Le et al., 2019), or by using queries generated by people from across the political spectrum (Mustafaraj et al., 2020; Trielli & Diakopoulos, 2020).

### *Limitations*

There are several important limitations to our study. First, results collected using a different selection of queries, search origins, and query days will likely differ from the ones we found. Second, the labeling of the results was performed via examination of the webpages of the location results, as well as whatever other information Google was able to provide, without calling or visiting the physical locations (c.f. (Bryant & Levi, 2012)). In that sense, the database provided by ANSIRH (see Methods) may be a more accurate "view" of existing abortion services, even though we show a correlation with their data at the state level (Spearman's $\rho$ = .624). Third, Google is a black box, and any assertions about its algorithm in this paper are speculation, such as the changes we observed that may be due to the BERT update (Nayak, 2019). A collaboration with the search engine would be invaluable toward understanding why we observe certain patterns, such as identifying distance constraints on location results that disadvantage those living in low-resource areas.

Fourth, there are other methods people seeking abortion services may employ to find an appropriate clinic or hospital – they may simply call the closest major hospital, ask advice from friends, or hear about a location in the news – and our method does not examine these alternatives. Also, we examined only the location results in this study, not the generic website results or ads, which can compose a substantial proportion of the page and provide different types of information that would likely affect a real searcher. Finally, perhaps due to the sensitive nature of the topic, it has been difficult to obtain data on existing locations providing abortion services, and our ground truth is limited to the preexisting databases we used. We urge the research community to set up a data sharing arrangement such that availability to health services can be monitored without exposing abortion clinics to potential harassment or abuse.

## Conclusion

The results of this study emphasize the necessity of monitoring the performance of technological tools which are becoming integral to the well-being of society. We illustrate that, while Google Search returns location results to actual abortion clinics in most cases, the number of results is unevenly distributed across locales. More focus on the quality of results and their coverage should be encouraged, for instance, by returning abortion clinic results which are further away but still relevant. We urge the research community to continue examining the fairness and equality of information access, and policymakers to engage with both researchers and companies to improve accountability and transparency, especially when concerning vulnerable populations.

**Abbreviations**

AC: abortion clinic(s)
ANSIRH: Advancing New Standards in Reproductive Health
CPC: crisis pregnancy center(s)
SERP: search engine results page(s)
TRAP: legal restrictions on abortion providers

**Acknowledgements**

We thank the Lazer Lab, Christo Wilson, and Piotr Sapieżyński, for helpful comments and discussions on this work. We also thank ANSIRH for making their data available for this research. Northeastern University was given permission to query Google Search automatically for research purposes. We note that Google did not review our research design, nor had any review rights with respect to the manuscript.